# Steering laser-produced THz radiation in air with superluminal ionization fronts

Silin Fu[1,2], Baptiste Groussin[1], Yi Liu[3,4], André Mysyrowicz[1], Vladimir Tikhonchuk[5,6], Aurélien Houard[1,*]

[1]*Laboratoire d'Optique Appliquée, ENSTA Paris, Ecole Polytechnique, CNRS, Institut Polytechnique de Paris, 91762 Palaiseau, France*
[2]*School of Nuclear Science and Technology, Lanzhou University, Lanzhou 730000, China*
[3]*Shanghai Key Lab of Modern Optical System, University of Shanghai for Science and Technology, 516, Jungong Road, 200093 Shanghai, China*
[4]*CAS Center for Excellence in Ultra-intense Laser Science, Shanghai, 201800, China*
[5]*Centre Lasers Intenses et Applications, Université de Bordeaux-CNRS-CEA, 351 Cours de la Liberation, 33405 Talence cedex, France*
[6]*Extreme Light Infrastructure ERIC, ELI-Beamlines Facility, Za Radnicic 835, 25241 Dolní Břežany, Czech Republic*
* *Corresponding author: Email: aurelien.houard@ensta.fr*

**Abstract:** We demonstrate that pulsed THz radiation produced in air by a focused ultrashort laser pulse can be steered to large angles or even in the backward direction with respect to the laser propagation axis. The emission angle is adjusted by the flying focus technique, which determines the speed and direction of the ionization front created by the single-color laser pulse. This easily adjustable THz source, being well separated from the intense laser, opens exciting applications for remote THz spectroscopy.

Introduction. - The filamentary plasma column produced by a chirp-free single-color femtosecond laser pulse propagating in air acquires a heavily damped oscillation both in the transverse and longitudinal directions with respect to the laser propagation axis. With a typical laser intensity $I < 10^{16}$ W/cm$^2$ typical of air filaments, the longitudinal oscillation due to the Lorentz force is much smaller than the transverse one. It is responsible for the emission of a short THz radiation pulse peaked at the plasma oscillation frequency along a forward-oriented cone, with a conical angle $\theta$ scaling like $\theta = (\lambda/L)^{1/2}$ where $L$ is the length of the plasma string, and $\lambda$ is the THz wavelength. It corresponds to a coherent transition-Cherenkov radiation stemming from the interference between the ends of the plasma string propagating at the speed of light [1-7]. Using a pulse shaping technique known as the flying (or sliding) focus [8-14], it is now possible with a focused laser pulse to control the velocity and direction of the ionization front through an interplay of chromatic aberration and laser pulse phase chirping, allowing exciting new applications such as photon acceleration [15] or improved phase-matching in x-ray lasers [16].

In this article, we demonstrate that the use of the flying focus technique with a single-color laser pulse can produce THz radiation in air that can be tuned to large angles with respect to the laser propagation or even in the backward direction. This is in contrast with the use of the flying focus with a two-color femtosecond pulse, which limits the THz emission to narrow cones in the forward laser direction, as theoretically shown in Refs. [10,17]. It requires a simple experimental set-up consisting of a short laser pulse from a chirp-controlled laser system focused with a normal lens uncompensated for chromatic aberration. With superluminal ionization velocities $v > c$, the THz energy is increased by almost an order of magnitude over the transition-Cherenkov radiation obtained with $v = c$, and all THz frequencies are emitted along a narrow cone at large angle. The radially polarized THz emission corresponds then to a classical Cherenkov radiation, produced, not by a charged particle, but by a dipole moving in a dielectric medium at a velocity higher than the light velocity [18-20]. The ability of this THz source to emit backward or around 90° could be very useful for remote spectroscopy [21] or THz imaging [22] since the THz radiation is well separated from the intense laser pulse.

Experimental setup – The measurements were performed using a CPA (chirped pulsed amplified) laser system from Amplitude technologies focused in air to produce a millimeter length plasma channel. It can deliver laser pulses at 10 Hz with a maximum energy of 200 mJ, a minimum pulse duration of 50 fs, a central wavelength at 790 nm and a bandwidth of 20 nm FWHM. To control the phase velocity $v$ of the pump pulse at the focus, the pulse was chirped to a pulse duration $\tau_{\text{las}}$ by detuning the grating compressor of the laser. This chirped pulse was then focused using a plano convex refractive lens of focal length $f = 30$ cm with a calibrated chromatic aberration. The focal distance of the 780 nm and 800 nm components are 30.06 cm and 29.94 cm, respectively. The velocity of the flying focus could then be tuned by adjusting the input chirp of the laser [8-9]. In the experiment, we used an input energy of 11 mJ to generate ionization in air with pulse durations ranging from 50 fs to 8.7 ps. To monitor the velocity of the plasma ionization front, a probe beam was sent perpendicularly to the plasma string (see FIG. 1(a)). This probe beam was a sample of the amplified beam extracted after the last amplifier and sent to an external compressor. It had a fixed pulse duration of 50 fs and an energy of a few nJ. After passing through the plasma, the probe beam was imaged on a CMOS camera or a wavefront sensor to perform time-resolved diffractometry [23-24]. Interferometric imaging of the plasma channel was performed to measure the plasma density and radius. The THz emission from plasma was collected with a Teflon lens of 40 mm diameter and focused on a heterodyne detection system, including a Gunn diode working at a frequency of 0.09 THz with a bandwidth of 4 GHz. The collected signal was amplified by a low noise 25 dB broadband amplifier. The sensitivity of this detector with an amplifier was 0.5 V/fJ for a picosecond THz pulse. The THz detector was mounted on a rotating stage and placed 50 cm away from the plasma (see FIG. 1 (b)). The angular aperture of the heterodyne detector was 4.6°.

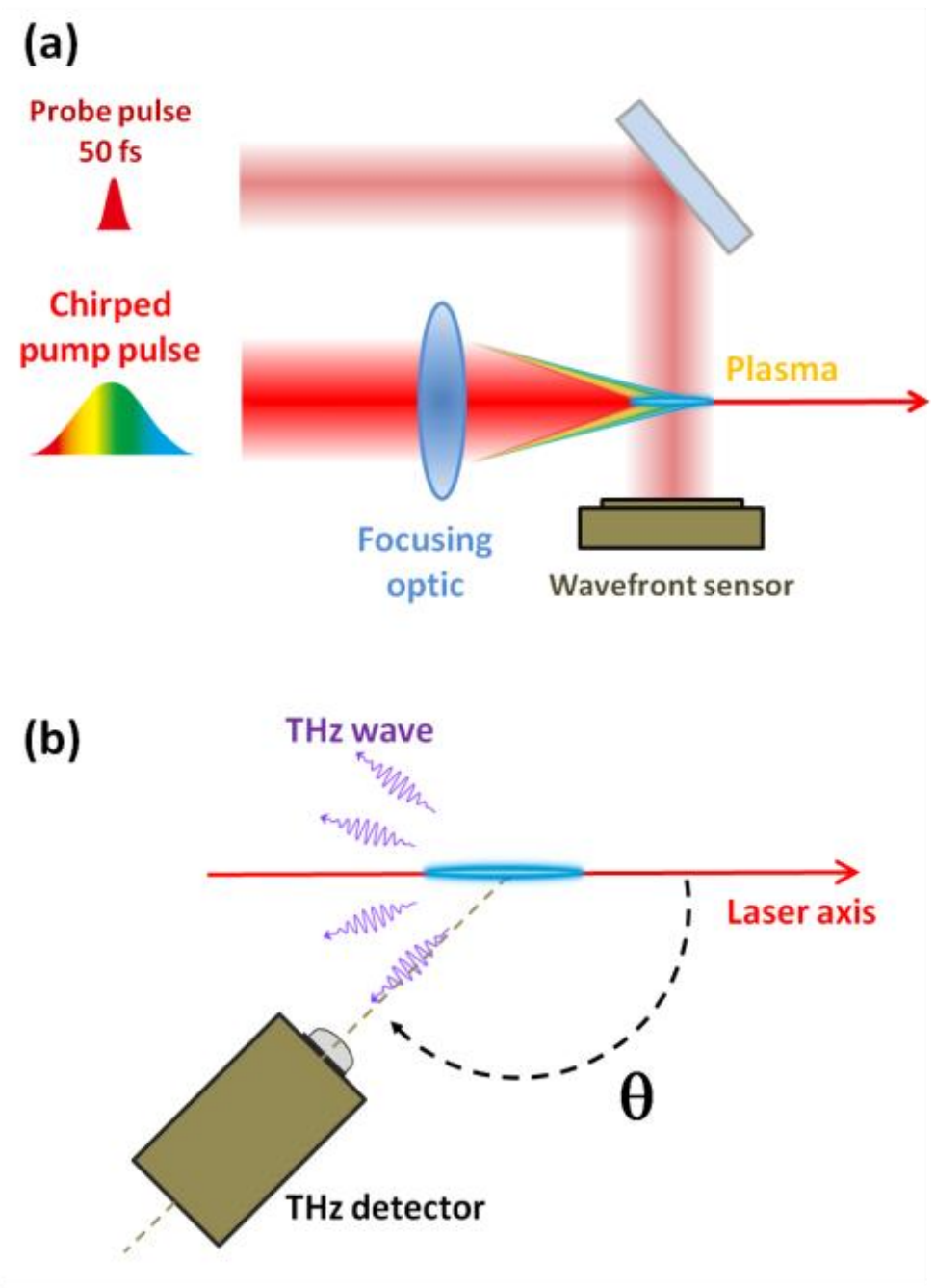

FIG. 1. Experimental setup. (a) A chirped laser pulse is focused in air with the flying focus technique to produce a plasma channel. A femtosecond pulse is used to probe the plasma length evolution as a function of time using transverse interferometry or diffractometry. (b) The radiation from the plasma string at 0.1 THz is measured as a function of angle $\theta$ using a heterodyne detector.

Results - We first characterized the displacement of the laser-induced ionization front with transverse diffractometry. An example of diffraction images obtained at delays of 0, 1.6, and 3.3 ps is shown in FIG. 2(a). One can see that the plasma is developing toward negative $z$, corresponding to a backward propagation. As shown in FIG. 2(b) the evolution of the ionization front was linear with the delay in all cases, allowing the extraction of a value of the ionization front velocity $v$ for each pulse duration. We observed a negative velocity for $\tau_{\mathrm{las}} = -8$ ps and $-6.4$ ps (the sign of $\tau_{\mathrm{las}}$ accounts for the sign of the laser chirp $C$), a velocity $v = c$ for the 50 fs pulse, and a superluminal positive velocity for $\tau = -2$ ps. These measurements are compared in FIG. 2(c) with the theoretical formula for the flying focus derived in Ref. [9]

$$\frac{v}{c} = \left(1 \pm \frac{c\tau_{\mathrm{las}}}{l}\right)^{-1}, \quad (1)$$

where $l$ is the distance between the blue and red pulse components produced by the lens at the focus, and the sign ± corresponds to the sign of the laser chirp. We measured a value of $l = 1.2$ mm for the lens. The theoretical curves for $v/c(\tau_{\mathrm{las}})$ are displayed as continuous lines in FIG. 2(c). We can see that the measurements are in very good agreement with the theoretical curves and that the velocity of the front is subluminal for positive chirp and superluminal for negative chirp.

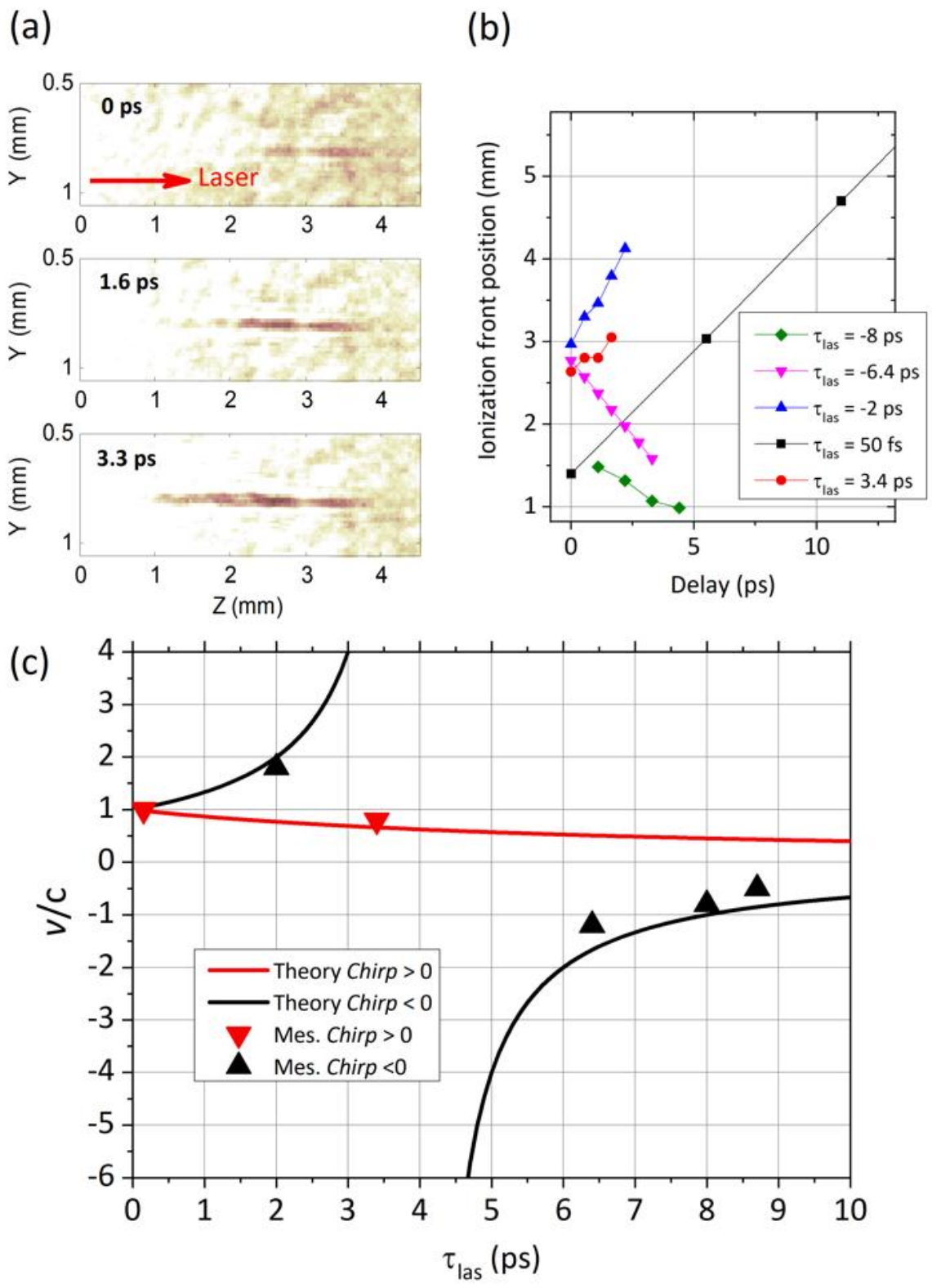

FIG. 2. Flying focus characterization. (a) Shadowgraphy images of the plasma for three delays for a negatively chirped pulse of 6.4 ps. (b) Measured position of the ionization front as a function of the delay $\Delta t$ for different pulse durations. (c) Ionization front velocities obtained from (b) for the different pulse durations. Solid lines show the theoretical value of $v$ based on the flying focus model [9]. Black lines correspond to negative chirp *C* and red line to positive chirp *C*.

We then characterized the radiation pattern at 0.1 THz from the plasma string for the different pulse durations considered in FIG. 2 corresponding to subluminal, luminal, and superluminal velocities of the ionization front *v*. The input energy of the laser was fixed to 11 mJ, and we measured an average plasma length of ~1 mm, except for the Fourier transform limited pulse that was two times longer. The results are presented in FIG. 3. In every case, we measured a conical THz emission. For the shortest pulse duration, the maximum emission is close to 20°, while for longer negative pulse durations, the angle of the emission approaches 90°. We note the presence of pronounced THz emission in the backward direction for negative velocities close to c. We have investigated the influence of the input laser energy on the THz emission by varying energy from 20 to 100 mJ. As shown in FIG. 3(c) and (d), the THz signal is strongly amplified and broadened while maintaining the same emission direction. The THz pulse energy measured at 0.1 THz increases almost linearly with the laser input energy. In the case of forward THz emission from a plasma generated by a 1 mJ fs laser pulse, the conversion efficiency for THz radiation has been measured to be about $10^{-9}$, producing a THz pulse in the pJ range [25-27]. This value is in good agreement with theoretical estimates [1,2] and with the measurement of our heterodyne detector at 0.1 THz. With a 100 mJ laser pulse, one can expect a THz energy of 10 to 100 pJ, depending of the laser pulse duration and focusing condition. As previously shown for the transition Cherenkov emission, this energy could be further increased by two orders of magnitude by applying a longitudinal static electric field [28-29]. Using an external DC field and an input laser energy of 100 mJ, one could expect to generate THz pulse with a nanojoule energy.

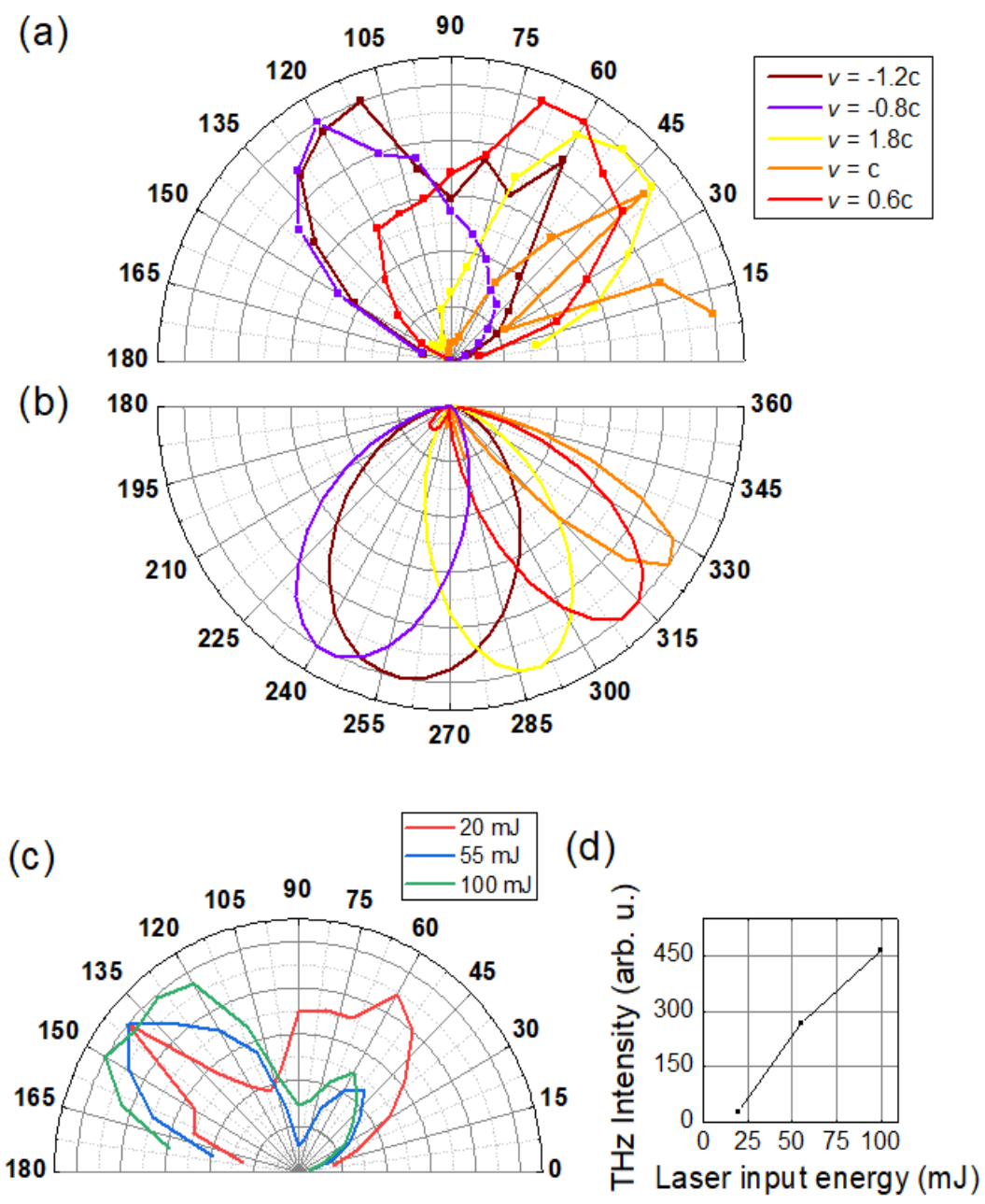

FIG. 3. (a) Measured THz angular distribution at 0.1 THz for different ionization front velocities $v$ and for an input pulse energy of 11 mJ. (b) Calculated angular distributions at 0.1 THz for the same ionization front velocities. (c) Measured THz angular distribution at 0.1 THz for a laser pulse with a duration $\tau = -8.7$ ps and an energy ranging from 11 to 100 mJ. All diagrams are normalized, and $\theta = 0$ corresponds to the laser propagation direction. (d) Backward THz intensity (above 90°) measured as a function of the laser input energy (same measurement as in panel (c)).

Theoretical model **-** A theoretical model of THz emission from a short laser pulse propagating in a plasma filament was developed in [2]. It includes two main elements: the evaluation of the amplitude and the temporal shape of the charge density distribution left behind the laser pulse and the evaluation of the electromagnetic emission from that moving charge propagating at the speed of light $c$. Since the plasma is neutral, the total electric charge in the wakefield is zero, but because of the different mobility of electrons and ions, there is a charge separation, which can be considered a flying dipole. In the theoretical model of Ref. [2], $v = \mathrm{c}$ and the true Cherenkov emission is forbidden, but electromagnetic emission is still produced because of the finite length of the plasma filament (for a detailed analysis of the relation between the Cherenkov radiation, transition, and bremsstrahlung radiation see also [5]). The theoretical approach can be generalized to a superluminal light velocity if the light velocity $c$ is replaced by the ionization front velocity $v$ given by Eq. (1). Consequently, equation (10) in [2] for the spectral intensity of electromagnetic emission in a solid angle $d\Omega$ also applies for the superluminal laser pulse and can be written as follows:

$$\frac{d^2W}{d\omega\, d\Omega} = \frac{q_{\mathrm{las}}^2\omega^2L^2}{4\epsilon_0c^3}\exp\left(-\frac{\omega^2\tau_{\mathrm{las}}^2}{2}\right)\frac{\omega^2(\omega^2+4\nu_e^2)}{\left(\omega^2-\omega_{\mathrm{pe}}^2\right)^2+\nu_e^2\omega^2}\,\mathrm{sinc}^2\left(\frac{\omega L}{2c}(\cos\theta - c/v)\right)\sin^2\theta. \quad (2)$$

Here $q_{\mathrm{las}} = er_{\mathrm{ch}}^2\omega_{\mathrm{pe}}^2\tau_{\mathrm{las}}I_{\mathrm{las}}/4m_evc\omega_{\mathrm{las}}^2$is the effective dipole charge created by a laser pulse with intensity $I_{\mathrm{las}}$, frequency $\omega_{\mathrm{las}}$, and duration $\tau_{\mathrm{las}}$; $\omega_{\mathrm{pe}}$ is the electron plasma frequency in the laser-created channel of radius $r_{\mathrm{ch}}$ and length $L$; $m_e$ and $e$ are the electron mass and charge, $\nu_e$ is the electron collision frequency, $\theta$ is the emission angle with respect to the laser propagation direction, and $\epsilon_0$ is the vacuum dielectric permittivity. The spectral distribution of the emission is controlled by two factors in this expression: the exponential factor represents the spectral distribution of laser pulse intensity, which is assumed to have a Gaussian temporal profile, and the Lorentzian term containing the plasma frequency and collision frequency, which describes the plasma response with a maximum of emission at the plasma frequency. Finally, the *sinc* function describes the Cherenkov effect with the cone emission angle defined by the resonance condition, $\cos\theta = c/v$. For $v = c$, we retrieve the transition-Cerenkov radiation observed in [2]. In this case, the maximum angle of emission is given by $\theta = (\lambda/L)^{1/2}$.

Using Eq. (2), we calculated the theoretical angular THz distributions for each measured case in FIG. 3. The plasma diameter, length, and density were measured using transverse interferometry [23-24]. We found $r_{\mathrm{ch}} = 50$ µm, a plasma density varying from $\rho = 3 \times 10^{16}$ cm$^{-3}$ to $10^{18}$ cm$^{-3}$, and a plasma length varying from $L = 0.8$ mm to 6 mm depending on the pulse duration. We considered a laser intensity in the filament of $5 \times 10^{13}$ W/cm$^2$ [30] and a collision frequency of $6 \times 10^{12}$ s$^{-1}$ [27]. The experimental results presented in FIG. 3(a) are compared to the calculated results in FIG. 3(b). The agreement is satisfactory, taking into account the fact that the plasma string is not as uniform as in a classical filament with intensity clamping and that the self-focusing effect perturbs the flying focus effect. Using this model, we calculated the evolution of the total THz radiated energy as a function of the ionization front velocity. We observe an increase of the THz radiation by a factor 5 between the transition-Cherenkov regime ($v = c$), and the Cherenkov regime ($v > c$). A similar increase is observed experimentally with the THz energy detected at 0.1 THz (see FIG. 4).

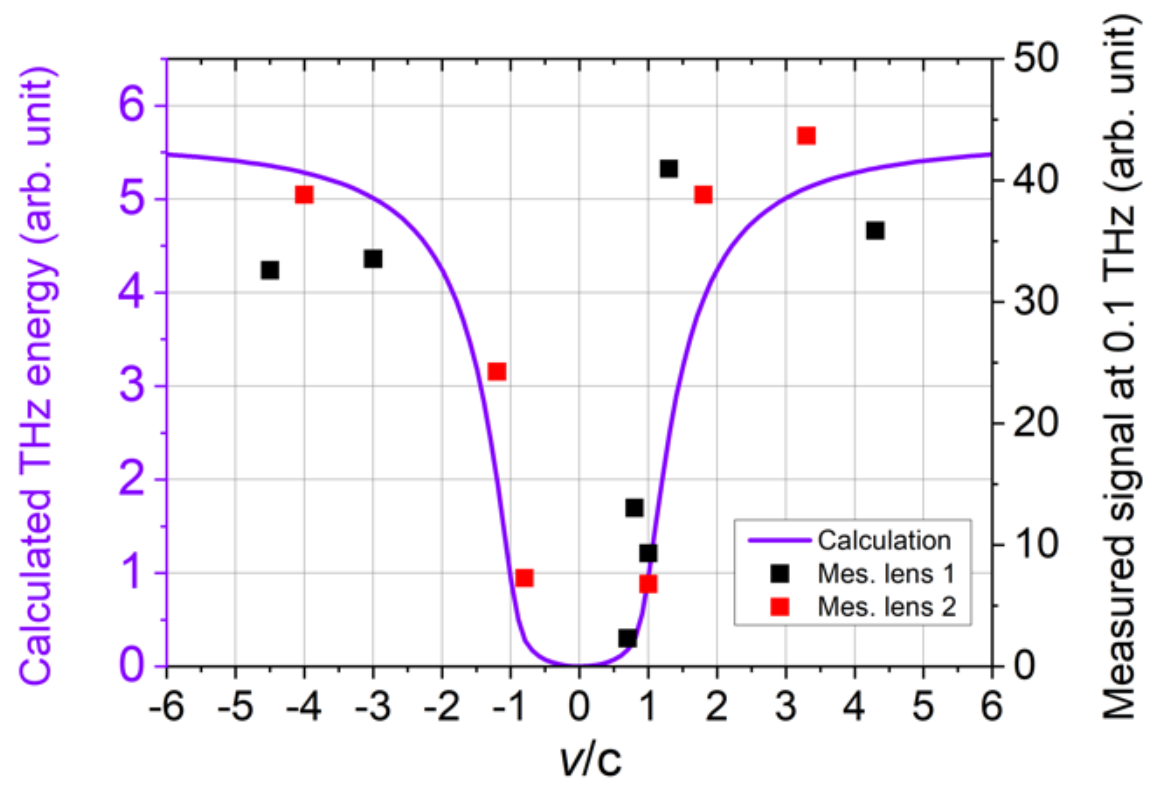

FIG. 4. Calculated THz energy as a function of the ionization front velocity (violet solid line) compared to the measured THz intensity at 0.1 THz obtained with two different lenses producing different aberrations. In this calculation, we consider a plasma length of 1 mm, with a density of $10^{16}$ cm$^{-3}$.

Finally, using formula (2), we calculated the angle for maximum THz emission as a function of $v/c$ (black curve in FIG. 5(a)) in the range [-6; +6]. For $v = c$, we retrieve the transition-Cerenkov radiation observed in [1]. In this case, the maximum angle of emission is given by $\Theta = (\lambda/L)^{1/2}$ and the energy is in the picojoule range with a 1 mJ laser pulse at 800 nm. When we decrease the velocity of the front below $c$, the angle of emission converges to 90°, and the radiated energy decreases. It corresponds then to a coherent transition radiation [31]. On the contrary, when $v > c$, the angle of emission progressively increases as the radiated energy. For $v > 1.5c$, the curve perfectly matches the theoretical angle for Cerenkov emission given by $\theta = \arccos(c/v)$ and plotted as a blue dashed line in FIG. 5(a). In the same figure, we also plot the barycenters of the THz angular distribution calculated using Eq. (2), with the measured plasma density and length for different ionization front velocities (red points) in FIG. 5(a). The black points correspond to the experimental measurements. It can be observed that the angles of THz emission predicted by the theory are in good agreement with the experimental results obtained for different flying focus velocities. Note that in the cases corresponding to positive velocities, the plasma length was about two times larger than in the cases corresponding to negative velocity, which explains why the curve is not symmetric with respect to $v/c = 0$. We also calculated the THz energy distribution as a function of angle $\Theta$ and frequency $\nu$, as shown in FIG. 5(b-e). Panel (b) corresponds to the subluminal case with $v = 0.8$c, (c) to the case with $v =$ c, (d) to the superluminal cases with $v = 5$c and (e) to a negative velocity $v = -1.7$c. The coalescence of all the THz frequencies on a narrow emission angle is clearly observed when the velocity is superluminal, indicating that this type of THz source could be advantageous for spectroscopy. At higher ionization front velocity, the angle of the emission tends to 90°. It is instructive to compare these results with those obtained with a flying focus in a two-color laser scheme. In that latter case, the THz radiation is maximized along the propagation axis [17, 32].

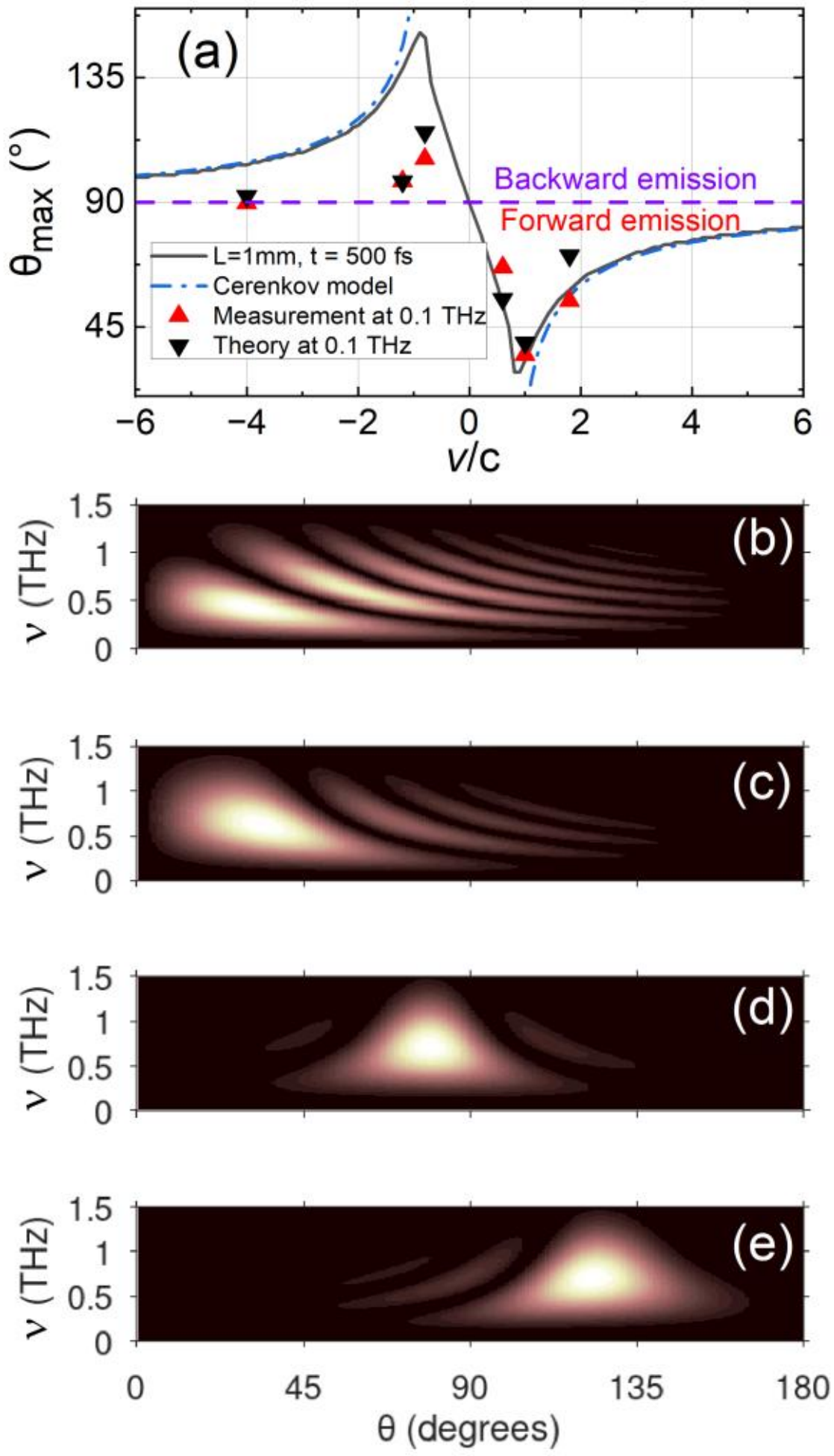

FIG. 5. (a) Calculated angle for maximum THz emission for the full spectrum (black continuous curve) and for a Cerenkov radiation (blue dashed curve) as a function of the ionization front velocity *v/c*. The red and black dots represent the calculated and measured barycenters of the THz angular distribution for the different ionization velocities shown in Fig. 3(a). (b-e) Calculated THz energy distribution emitted by a 1 mm plasma column as a function of angle $\theta$ and frequency $\nu$ for ionization velocities $v = 0.8c$, $c$, $5c$, and $-1.7c$.

Conclusion – We have demonstrated that THz radiation emitted by a short filament in air can be tuned to any direction. It requires a simple experimental set-up consisting of a focused short laser pulse from a chirp-controlled laser system and a chromatic aberrator. In contrast to the two-color THz generation scheme, there is no need to adjust an optical phase between two laser pulses. Forward and backward superluminal ionization fronts velocities $v$ of the generated plasma string can be achieved. With superluminal ionization velocities $v > c$, the THz energy is increased by almost an order of magnitude over the transition-Cherenkov radiation obtained with $v = c$, and all frequencies are emitted along a narrow cone at a large angle. We also observe that the Cherenkov THz yield increases linearly with the laser input energy while keeping the same angular emission. Application of a static electric field to the plasma could further enhance the THz emission by two orders of magnitude. THz emission could also be obtained in the laboratory at longer or shorter wavelengths [33] with the same flying focus technique by changing the plasma density using different gases.

## Acknowledgements

We acknowledge the technical support from Magali Lozano and Fatima Alahyane. We also thank Jean-Philippe Goddet and Cedric Thaury for helpful discussions.

This work has benefited from a grant managed by the Agence Nationale de la Recherche (ANR), as part of the program "Investissements d'Avenir" under the reference (ANR-18-EURE-0014) and from a grant by the National Natural Science Foundation of China (Grant No.12034013).